\documentstyle[psfig,12pt]{article}
\begin{document}
  
\centerline{\large\bf DARK MATTER HALOS IN A}
\vspace*{2mm}
\centerline{\large\bf SECONDARY INFALL MODEL}
  
\vspace*{0.5 cm}
 
\centerline{Antonino DEL POPOLO$^{(*)}$ \  and  \ Mario GAMBERA$^{(**)}$}
\footnotetext{$^{(*)}${\em Facolt\`{a} di Ingegeneria, Universit\`{a}
Statale di Bergamo, Dalmine (BG), Italy.}}
\footnotetext{$^{(**)}$ {\em Osservatorio Astrofisico di Catania and CNR-GNA, 
Viale A.Doria, 6 - I 95125 Catania, ITALY (e-mail: mga@sunct.ct.astro.it)}}
\vspace*{0.1 cm}
{\small
\begin{center}
{\em Istituto di Astronomia dell'Universit\`a di Catania, Viale A.Doria 6,\\
 I 95125 Catania, ITALY}
\end{center}
}
\vspace*{0.3 cm}

\vspace*{0.8 cm}
\begin{center}
{\bf ABSTRACT}
\end{center}
 
{\small
~\\ 
We calculate the density profiles of virialized halos in the
case of structure evolving hierarchically from a scale-free Gaussian $\delta
-$field having a power spectrum $P(k)\propto k^n$ in a $\Omega=1$ Universe; we suppose that the initial density contrast profile around
local maxima is given by the mean peak profile introduced by Bardeen et al.
(1986 hereafter BBKS).
We show both that the density profiles are not power-laws but have
a logarithmic slope that increases from the inner halo to its outer parts
and for $n \geq -1$ are well approximated by
Navarro et al. (1995, 1996, 1997) profile and the radius $a$, at which the slope $\alpha=-2$, is a function
of the mass of the halo and of the spectral index $n$. 
}
~\\

\section{Introduction}

The collapse of perturbations onto local density maxima of the primordial
density field is likely to have played a key role in the formation of
galaxies and clusters of galaxies. The problem of the collapse
has been investigated from several authors (Gunn \& Gott 1972; Gunn 1977;
Kaiser 1984; Davis et al. 1985; Hoffman \& Shaham 1985 - hereafter HS; BBKS; 
Hoffman 1988; Efstathiou et al. 1988; Evrard
et al. 1993; Navarro et al. 1995, 1996, 1997;
Avila-Reese et al. 1998).\\
To overcome the problem of the excessively steep density profiles,
$\rho\propto r^{-4}$, obtained
in numerical experiments of simple gravitational collapse Gunn \& Gott (1972),
Gott (1975) and Gunn (1977) were able to produce shallower profiles, 
$\rho\propto r^{-2}$ through the ${\it secondary}$ ${\it infall}$ process. 
HS considered a scale-free initial
perturbation spectra, $P(k) \propto k^n$ and assumed that
local density extrema are the progenitors of cosmic structures and that the
density contrast profile around maxima is proportional to
the two-point correlation function. They thus showed that $\rho\propto {r^{-\alpha}}$ with
$\alpha=\frac{3(3+n)}{(4+n)}$, thus recovering Bertschinger's (1985) profile for $n=0$ and
$\Omega=1$. They also showed that, in an open Universe, the slopes of
the density profiles steepen with increasing values of $n$ and
with decreasing $\Omega$, reaching a profile
$\rho\propto r^{-4}$ for $\Omega \rightarrow 0$. Hoffman (1988) refined the
calculations by HS and made a detailed comparison
of the analytical predictions of the {\it Secondary Infall Model} 
(hereafter SIM) with the simulations by Quinn et al. (1986) and 
Quinn \& Zurek (1988). 
The good results
given by the SIM in describing the formation of dark matter halos seem to be
due to the fact that in energy space the collapse is ordered and gentle,
differently from what seen in N-body simulations (Zaroubi et al. 1996).\\
A great effort has been dedicated to study
the role of initial conditions in shaping the final structure of the dark
matter halos; but, if on large scales
(evolution in the weakly non-linear regime)
the growing mode of the initial density fluctuations can be recovered if the
present velocity or density field is given
(Peebles 1989; Nusser \& Dekel 1992), on small scales shell crossing and
virialization contribute to make the situation less clear. 
To study the problem,
three-dimensional large-scale structure simulations were run with 
often conflicting results (Quinn et al. 1986; West et al. 1987; Efstathiou et al. 1988). 
More recent studies (Voglis et al. 1995; Zaroubi et al. 1996) showed a
correlation between the profiles and the final structures. Besides Lemson (1995), Cole \& Lacey (1996), 
Navarro et al. (1996, 1997) and Moore et al. (1997) found that dark
matter halos do not follow a power law but
develop universal profile, a quite general profile
 for any scenario in which structures form due to hierarchical
clustering, characterized by a slope $\beta=\frac{d \ln \rho}{d ln r}=-1$ near
the halo center and $\beta=-3$ at large radii.
In that approach, density profiles can be fitted with a
one parameter functional form:
\begin{equation}
\frac{\rho(r)}{\rho_b}= \frac{\delta_n}{\frac{r}{a}\left(1+\frac{r}{a}\right)^2}
\label{eq:nfw}
\end{equation}
where $ {\rho_b}$ is the background density and $ {\delta_n}$ is the 
overdensity [below we shall refer to Eq. (\ref{eq:nfw}) (Navarro et al. 1997) 
as the NFW profile].
The scale radius $a$, which defines the scale where the
profile shape changes from slope $\beta<-2$ to $\beta>-2$,
and the characteristic overdensity, $\delta_n$, are related because the
mean overdensity enclosed within the virial radius $r_{vir}$ is $ \simeq 180$.
The scale radius and the central overdensity are directly related to the
formation time of a given halo (Navarro et al. 1997). The power spectrum
and the cosmological parameters only enter to determine the typical formation
epoch of a halo of a given mass, and thereby the dependence of the
characteristic radius on the total mass of the halo. Also these last results
are not universally accepted. In short, the question of whether galaxies and clusters
mass density profiles retain information on the initial conditions and
the evolutionary history that led to their formation
remains an open question. \\
In this paper, we introduce a modified version by
HS and Hoffman's (1988) models to study the shapes 
of the density profiles that result from the gravitational collapse. 
In particular, we relax the hypothesis that the initial density profile 
is proportional to the two-point correlation function, and use the density 
profiles given by BBKS.  
The plan of the paper is the following: in Sect. ~2 
we introduce our model and in Sect. ~3 we show our results and we draw our conclusions.

\section{The model}

In the most promising cosmological scenarios, structure formation 
in the universe is generated through the growth and collapse 
of primeval density 
perturbations originated from quantum fluctuations (Guth \& Pi 1982; 
Hawking 1982; Starobinsky 1982; BBKS) in an inflationary 
phase of early Universe. 
The growth in time of small 
perturbations is due to gravitational instability. 
The statistics of 
density fluctuations originated in the inflationary era are Gaussian, and 
can be expressed entirely 
in terms of the power spectrum of the density fluctuation sP( k). 
In biased structure formation theory it is assumed that cosmic structures 
of linear scale $ R_f$ form around the peaks of the density field, 
$  \delta( r)$, smoothed on the same scale. \\
If we suppose we are sitting on a $\nu \sigma$ extremum in the the smoothed
density field, we have that:
\begin{equation}
\delta(0)= \nu \xi(0)^{1/2} = \nu \sigma
\end{equation}
together with:
\begin{equation}
{\bf \bigtriangledown} \delta(r) \mid_{r=0}= 0
\end{equation}
where $\xi (r)$ is the two-point correlation function given in $ r = 0$
If the Laplacian of $\delta(r)$ is unspecified, that means that the extremum
may be a maximum or a minimum, 
the mean density at a distance $r$ from the peak is then:
\begin{equation}
\delta(r)= \nu \xi(r)/\xi(0)^{1/2}
\label{eq:delt}
\end{equation}
(Peebles 1984; HS). If we calculate the mean density
around maxima, as done by BBKS, by adding the constraint: 
\begin{equation}
{\bf \bigtriangledown}^2 \delta(r) \mid_{r=0}<0
\end{equation}
we find that the mean density around a peak is given by:
\begin{equation}
\langle \delta (r) \rangle =\frac{\nu \xi (r)}{\xi (0)^{1/2}}-\frac{\vartheta (\nu
\gamma ,\gamma )}{\gamma (1-\gamma ^2)}\left[ \gamma ^2\xi (r)+\frac{%
R_{\ast }^2}3\nabla ^2\xi(r) \right] \cdot \xi (0)^{-1/2} 
\label{eq:dens}
\end{equation}
(BBKS; Ryden \& Gunn 1987),
where $\nu $ is the height of a density peak, 
$\gamma $ and $R_{\ast}$ are two spectral parameters
while $ \vartheta (\gamma \nu ,\gamma )$ is: 
\begin{equation}
\theta (\nu \gamma ,\gamma )=\frac{3(1-\gamma ^2)+\left( 1.216-0.9\gamma
^4\right) \exp \left[ -\left( \frac \gamma 2\right) \left( \frac{\nu \gamma }%
2\right) ^2\right] }{\left[ 3\left( 1-\gamma ^2\right) +0.45+\left( \frac{%
\nu \gamma }2\right) ^2\right] ^{1/2}+\frac{\nu \gamma }2}
\label{eq:tet}
\end{equation}
In order to calculate $\delta(r)$ we need a power spectrum, $P(k)$.
In the following, we restrict our study to an Einstein-De Sitter ($\Omega=1$)
Universe with zero cosmological constant and scale-free density
perturbation spectrum $P(k)$
\begin{equation}
P(k)=A k^n
\end{equation}
with a spectral index in the range $-1 \leq n \leq 0$,
and also to a CDM Universe with 
spectrum given by BBKS.
We normalized the spectrum by  
imposing that the mass variance at $8h^{-1}Mpc$ is $\sigma _{8}=0.63$. 
In the case of a free-scale power spectrum, it
is easy to show that the two-point correlation
function can be expressed in terms of the confluent
hypergeometric function, $_1F_1$, and of the $\Gamma$ (see 
Del Popolo et al. 1999) function. 
In the case that $\nu$ is very large Eq. (\ref{eq:tet}) reduces to
\begin{equation}
\theta \rightarrow 3(1-\gamma^2)/(\nu \gamma)
\end{equation}
and the mean density is well approximated by Eq. (\ref{eq:delt}), which is
the approximation used by HS
to calculate $\delta(r)$. 
In reality, for peaks having $\nu=2,3,4$, the mean expected density profile
is different from
a profile proportional to the correlation function both for galaxies
and clusters of galaxies (see BBKS). For example for galaxies the CDM profile is
steeper than that proportional to $\xi(r)$ as shown by Ryden \& Gunn (1987)
with a discrepancy increasing with decreasing $\nu$. 
As shown by Gunn \& Gott (1972), a bound mass shell having initial comoving
radius $x$
will expand to a maximum radius:
\begin{equation}
r_m=x/{\overline \delta(r)}
\label{eq:pee}
\end{equation}
where
the mean fractional density excess inside the shell, as measured at
current epoch $t_0$, assuming linear growth, 
can be calculated as:
\begin{equation}
{\overline \delta}=\frac{3}{r^3} \int_0^r \delta(y)y^2 dy
\end{equation}
At initial time $t_i$ and for a Universe with density parameter $\Omega_i$,
a more general form of
Eq. (\ref{eq:pee}) (Peebles 1980) is :
\begin{equation}
r_m=r_i\frac{1+{\overline \delta_i}}{{\overline \delta_i}-(\Omega_i^{-1}-1)}
\end{equation}
The last equation must be regarded as the main essence of the SIM. It tells
us that the final time averaged radius of a given Lagrangian shell
does scale with its initial radius.
Expressing the scaling of the final radius, $r$, with the initial one
by relating $r$ to the turn around radius, $r_m$, it is possible to write:
\begin{equation}
r=Fr_m
\end{equation}
where $F$ is a costant that depends on $\alpha$.
(Zaroubi et al. 1996).
If energy is conserved, 
then the shape of the density profile at maximum of expansion is
conserved after the virialization,
and
is given by (Peebles 1980; HS; White \& Zaritsky 1992):
\begin{equation}
\rho(r)=\rho_i \left( \frac{r_i}{r} \right)^2 \frac{d r_i}{dr}
\end{equation}
In the limit $\nu>>1$, the overdensity $\delta(r)$ is proportional
to the two-point correlation function and the density profile is
a function of $n$ and $\Omega$ only, and then the expected profile is
that by HS. 

\section{Results and conclusions}

By using the model introduced in the previous section we have studied the
density profiles of halos in free-scale universes with
$-1 \leq n \leq 0$, and for a CDM model characterized by a BBKS spectrum. As
previously quoted, the chosen range of $n$ is dictated by the limits of
the SIM and by the values of $n$ interesting in the cosmological
context. \\
In Fig. 1 we plot the slope of the density profile for several
values of $n$ and $\nu$.
We began by finding the HS solution, that was recovered as expected
in the limit
$\nu>>1$. 
This solution is represented by the short-dashed line which 
coincides with the result by HS, namely 
$\alpha=\frac{3(3+n)}{(4+n)}$, 
indicating an increase in the slope $\alpha$ with increasing $n$.
\begin{figure}[ht]
\psfig{file=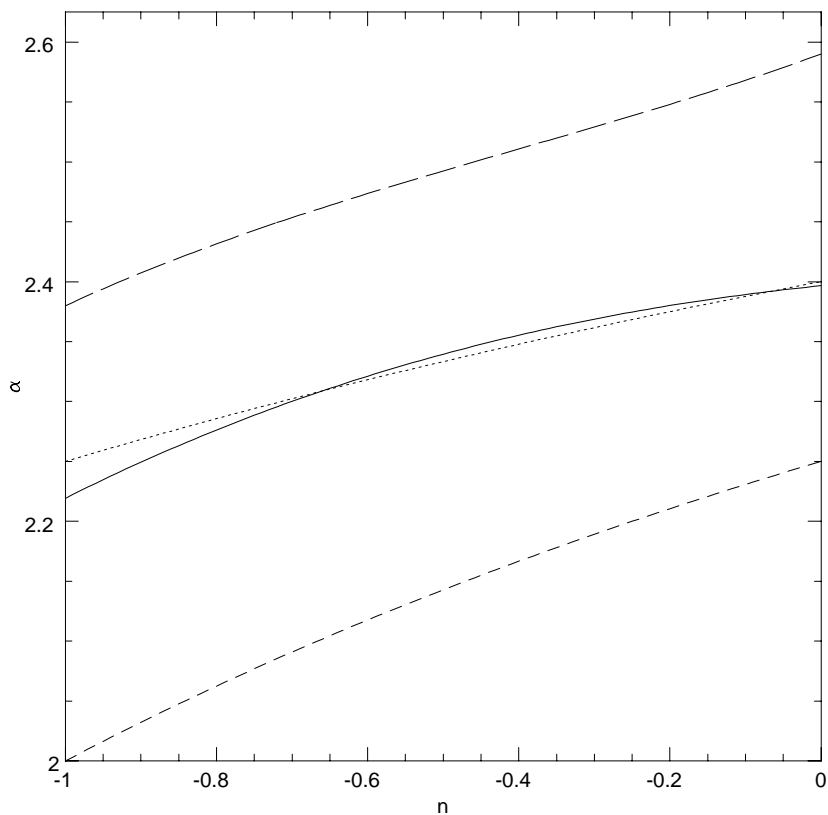,width=12.0cm}
\caption[]{The slope $\alpha$ of density profiles as a function of the spectral
index $n$ and $\nu$. The short-dashed line represents $\alpha$ in the limit $n>>1$.
It 
coincides with the HS result. The solid line represents the logarithmic slope
for $\nu=3$, while the dotted line is Shet \& Jain's (1996) result. The
long-dashed line represents $\alpha$ for $\nu=2$.}
\end{figure}
Because of the rarity of extremely high peaks, most galaxies and clusters
will form from peaks of height 2 or 3 $\sigma$ (BBKS; Ryden \& Gunn 1987): so
we repeated the calculation of $\alpha$ for these values. 
For $\nu=3$, the logarithmic slope of the density profile, calculated
at $1 h^{-1} Mpc$ 
is steeper for all values of
$n$ (solid line) than that obtained by HS, and it is well approximated
by Shet \& Jain (1996) (dotted line), $\alpha=\frac{3(4+n)}{(5+n)}$,
obtained using
stable clustering and neglecting halo-halo correlations. At the same time
the dependence of $\alpha$ on $n$ is weaker than that shown by HS.
\begin{figure}[ht]
\psfig{file=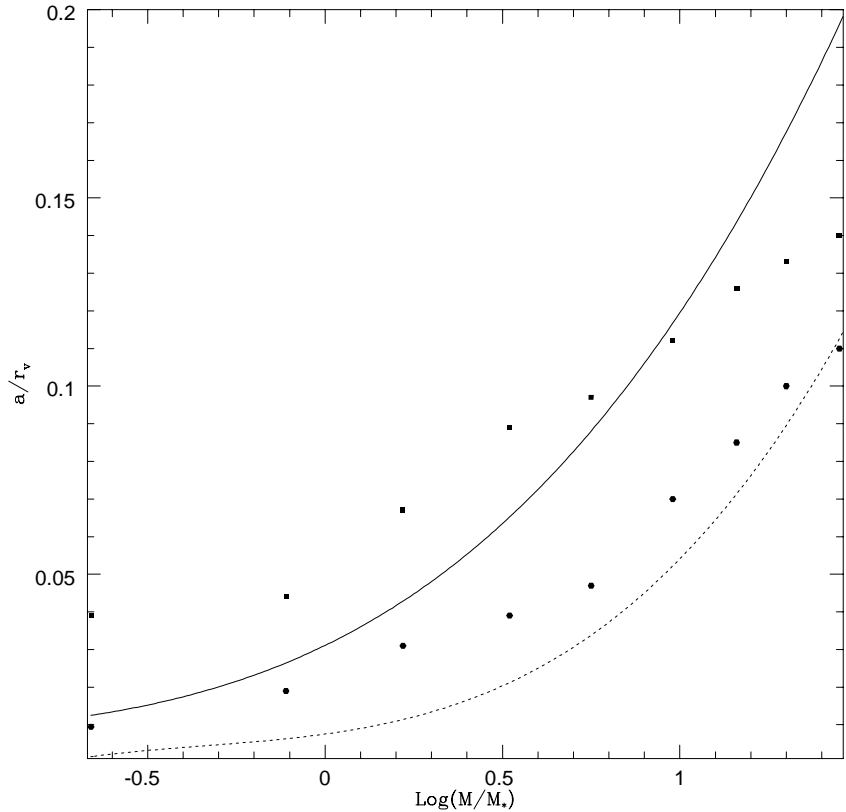,width=12.0cm}
\caption[]{Trend of the scale radius $a$ versus the mass of the halos in
the case $n=-1$ (solid line) and $n=0$ (dotted line).
The filled squares and the filled exagons
represents $a/r_v$, for $n=-1$ and $n=0$ respectively, obtained
by Navarro et al. (1997) in the case $f=0.01$.
}
\end{figure}
For $\nu=2$ the slope is even steeper than the previous case (long-dashed line)
and it is well approximated by Crone's et al. (1994) result, which
is also consistent with the results by Navarro et al. (1997) (see their Fig. 13). As
in the previous case the dependence of $\alpha$ on $n$ is weaker with
respect to that shown in HS.\\
In Fig. 2 we plot the variation of the scale parameters $a$ versus 
$M/M_{\ast}$. Masses are normalized by the characteristic mass $M_{\ast}$,
which is defined at a time $t$ as the linear mass on the scale
currently reaching the non-linear regime:
\begin{equation}
M_{\ast}(t)=\frac{4 \pi}{3}R_{\ast}^3 \rho_b(t)
\end{equation}
where the scale $R_{\ast}$ is such that the linear density contrast on this
scale is $\delta(R_{\ast})=1.69$ and $\rho_b(t)$ is the background density 
at the time $ t$. Once known that the mass variance,
$\sigma_M$,
for a power
spectrum $P(k) \propto k^n$ is given by $\sigma_M \propto R^{-(3+n)}$
and remembering our normalization $\sigma_M(8 h^{-1} Mpc)=0.63$, the value of
$M_{\ast}$ for $n=-1$ results to be $M_{\ast}=6 \times 10^{13} M_{\odot}$.
The value of the dimensionless scale radius $a$ correlates strongly
with halo mass and with spectral index $n$.
The
solid line represents $a$ for $n=-1$. As shown in Fig. 2,
more massive halos have
a larger scale radius $a$, or equivalently less massive
halos are more concentrated. The dotted line shows $a$ for $n=0$.
Also for this value of $n$ more massive halos are less centrally
concentrated. Finally from Fig. 2 we also see that
in models with more small-scale power (or equivalently
larger values of $n$) the haloes tend to have denser cores.
These results were expected because halos with mass $M<<M_{\ast}$
form much earlier than haloes with $M>>M_{\ast}$ and then are more
centrally concentrated. Moreover, for a fixed value of $M/M_{\ast}$,
haloes form earlier in models with larger values of $n$ and then have
denser cores. This result is in qualitative agreement with those by 
Navarro et al. (1997),
Cole \& Lacey (1996),  Tormen et al. (1997).
The filled squares and the filled exagons, in Fig. 2,
represents $a/r_v$ for $n=-1$ and $n=0$ respectively obtained
by Navarro et al. (1997) in the case $f=0.01$ (see their paper for
a definition of this parameter) which give the best
fit to the results of their simulations.\\
The virial radius $r_v$ is
obtained by using Navarro et al. (1997) equation and it 
determines the mass of the halo through:
\begin{equation}
M_v=200 \rho_c \frac{4 \pi}{3} r_v^{3}
\end{equation}
In the case $n=-1$, our model gives less concentrated halos till
$M \simeq 10 M_{\ast}$ and after this value the tendence is reversed.
In the case $n=0$, our model gives halos sligthly more concentrated
in the overall studied mass range. \\
The result
obtained is remarkable. Our model is based on spherical simmetry, and as we
previously stressed, halos accretion does not happen in
spherical shells but by aggregation of subclumps of matter which
have already collapsed. In other words it seems that the halos structures
does not depend crucially on hierarchical merging, in agreement with
Huss et al. (1998). The SIM seems to have more predictive
power than that till now conferred to it. 

\begin{flushleft}
{\it Acknowledgements}
\end{flushleft}
We are grateful to Proff. E. Recami and E. Spedicato for 
stimulating discussions during the period in which this work was performed.\\

%
\end{document}